\documentclass[preprint,preprintnumbers,amsmath,amssymb,prl,superscriptaddress]{revtex4}

\usepackage{graphicx}
\usepackage{dcolumn}
\usepackage{bm}

\begin{document}

\title{Andreev reflection at the interface with an oxide in the quantum Hall regime}

\author{Yusuke Kozuka}
\altaffiliation{KOZUKA.Yusuke@nims.go.jp; Present address: Research Center for Magnetic and Spintronic Materials, National Institute for Materials Science (NIMS), 1-2-1 Sengen, Tsukuba 305-0047, Japan}
\affiliation{Department of Applied Physics and Quantum-Phase Electronics Center (QPEC), University of Tokyo, Tokyo 113-8656, Japan}
\affiliation{JST, PRESTO, Kawaguchi, Saitama 332-0012, Japan}

\author{Atsushi Sakaguchi}
\affiliation{Department of Applied Physics and Quantum-Phase Electronics Center (QPEC), University of Tokyo, Tokyo 113-8656, Japan}

\author{Joseph Falson}
\affiliation{Max Planck Institute for Solid State Research, Heisenbergstrasse 1, D-70569 Stuttgart, Germany}

\author{Atsushi Tsukazaki}
\affiliation{Institute for Materials Research, Tohoku University, Sendai 980-8577, Japan}

\author{Masashi Kawasaki}
\affiliation{Department of Applied Physics and Quantum-Phase Electronics Center (QPEC), University of Tokyo, Tokyo 113-8656, Japan}
\affiliation{RIKEN Center for Emergent Matter Science (CEMS), Wako 351-0198, Japan}

\begin{abstract}
Quantum Hall/superconductor junctions have been an attractive topic as the two macroscopically quantum states join at the interface. Despite longstanding efforts, however, experimental understanding of this system has not been settled yet. One of the reasons is that most semiconductors hosting high-mobility two-dimensional electron systems (2DES) usually form Schottky barriers at the metal contacts, preventing efficient proximity between the quantum Hall edge states and Cooper pairs. Only recently have relatively transparent 2DES/superconductor junctions been investigated in graphene. In this study, we propose another material system for investigating 2DES/superconductor junctions, that is ZnO-based heterostrcuture. Due to the ionic nature of ZnO, a Schottky barrier is not effectively formed at the contact with a superconductor MoGe, as evidenced by the appearance of Andreev reflection at low temperatures. With applying magnetic field, while clear quantum Hall effect is observed for ZnO 2DES, conductance across the junction oscillates with the filling factor of the quantum Hall states. We find that Andreev reflection is suppressed in the well developed quantum Hall regimes, which we interpret as a result of equal probabilities of normal and Andreev reflections as a result of multiple Andreev reflection at the 2DES/superconductor interface.
\end{abstract}

\maketitle

\section{Introduction}

Andreev reflection is a phenomenon at the metal/superconductor interface where an electron with an energy less than the superconducting gap ($\Delta$) in the metal incident on the superconductor is reflected as a hole tracing the same trajectory of the incident electron (retroreflection) through creation of a Cooper pair in the superconductor as shown in Fig. 1(a) \cite{Andreev}. Upon Andreev reflection, rigid momentum and spin relationships are maintained between the incident electron and the retroreflected hole \cite{Blonder1}. This property has been utilized as a unique probe to detect the degree of spin polarization of the metal from zero-bias conductance enhancement or reduction, since Andreev reflection is forbidden for a spin-polarized ferromagnet/singlet-superconductor interface due to the absence of spin density of states for retroreflected holes [Fig. 1(b)] \cite{deJong,Soulen,Mazin}. Superconducting symmetry has also been deduced for $d$-wave cuprate superconductors based on the unusual shape of the Andreev spectra \cite{Deutscher}.\par

\begin{figure}[tbp]
  \begin{center}
    \includegraphics{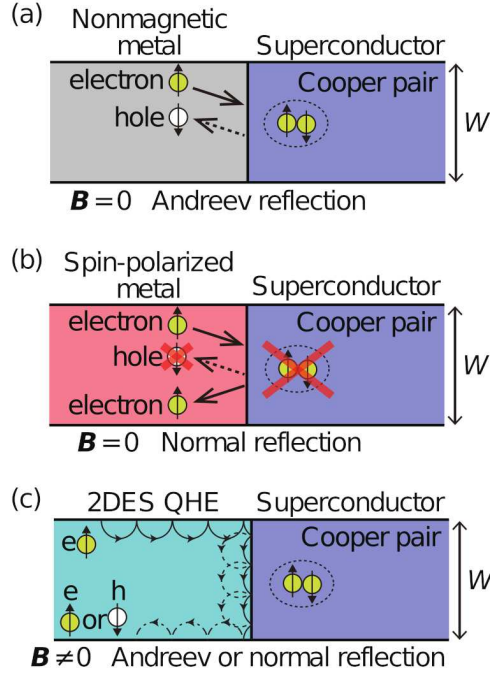}
  \end{center}
  \caption{(color online). Schematics of (a) Andreev reflection at the nonmagnetic metal/superconductor interface, (b) normal reflection at the spin-polarized ferromagnet/superconductor interface, and (c) 2DES/superconductor interface in the quantum Hall effect (QHE). In the case of (b), Andreev reflection is forbidden due to the absence of spin-down hole density of  states. In the quantum Hall regime of 2DES in (c), Andreev reflection is feasible through multiple Andreev and normal reflections at the 2DES/superconductor interface.}
\end{figure}

In the presence of a magnetic field ($B$), the aforementioned picture of Andreev reflection needs reconsideration because the reflected hole no longer traces back to the trajectory of the incident electron due to the Lorentz force. Particularly in the case of a two-dimensional electron system (2DES), the chiral edge modes in the quantum Hall states at first sight are not compatible with the Andreev reflection because of the absence of counterpropagating edge modes for retroreflected holes. However, in a schematic of skipping orbitals at the edge of the sample shown in Fig. 1(c), an electron incident on an edge of the 2DES/superconductor interface eventually comes out from the other edge of the interface either as an electron or a hole through multiple Andreev reflections (MAR) as well as normal reflections along the interface. The issue of Andreev reflection under quantizing magnetic field has been addressed in a number of theoretical \cite{Ma,Zyuzin,Fisher,Hoppe,Giazotto,Stone,Takagaki,Ostaay,Gunawardana} and experimental \cite{Takayanagi1,Moore,Uhlisch,Takayanagi2,Eroms,Matsuo,Batov,Komatsu,Rickhaus,Wan,Calado,Amet,Lee,Park,Kumaravadivel} studies for two-dimensional electron systems realized in semiconductor heterostructures or in graphene. Many of these studies indicated that Andreev reflection indeed coexist with the chiral edge modes albeit the detailed quantum mechanical understanding is not yet settled.\par

Recently, the interaction between superconductivity and chiral or helical edge states has attracted renewed attention because non-abelian quasiparticles are predicted to emerge as electron-hole hybrids at the interface \cite{Fu,Tanaka,Qi,Lindner,Clarke,Mong,Vaezi}. However, making good proximity between Cooper pairs and 2DES are not easily achieved due to Schottky barrier formation at the semiconductor surface \cite{Tung}. This problem originates from Fermi level pinning caused by dangling bond states at the semiconductor surface. In order to avoid this issue, InAs 2DES have frequently been employed since the pinning level tends to be formed outside of the band gap. However, the inherently high carrier density in InAs prevents the investigation of the quantum Hall effect at low filling factors ($\nu$) \cite{Takayanagi1,Eroms,Matsuo}. Only recently have high-quality superconductor/quantum Hall edge junctions been realized using graphene, free from the Schottky barrier problem \cite{Komatsu,Rickhaus,Calado,Amet,Lee,Park,Kumaravadivel}.\par

Here we utilize ZnO-based heterostructures to investigate Andreev reflection in the quantum Hall regime. Compared with covalent semiconductors, the ionic nature of oxide semiconductors usually suppresses the Fermi level pinning effect owing to less dangling bond states at the surface \cite{Kurtin}. Therefore, ideal ohmic contacts for 2DES can be easily formed with  low-workfunction metals such as Ti  \cite{Brillson,Kozuka} unlike the above-mentioned covalent semiconductors \cite{Tung}. Among oxide semiconductors, the 2DES in ZnO heterostructures is an exceptional candidate, exhibiting pronounced integer and fractional quantum Hall effects \cite{Tsukazaki1,Tsukazaki2,Falson1,Falson2,Falson3}. \par

In this study, we have fabricated 2DES/superconductor junctions on Mg$_{x}$Zn$_{1-x}$O/ZnO heterostructures, and investigated the Andreev reflection under magnetic field up to the quantum Hall regime close to $\nu=1$. MoGe is employed as a superconductor because of its high upper critical field in the amorphous form \cite{Grayl,Bezryadin}. At low temperature, the junction shows signatures of Andreev reflection, indicating a modest energy barrier at the interface. Under quantizing magnetic fields above 1 T, the Landau quantization causes a periodic modulation in the conductance, suggestive of the variation of the Andreev reflection probability. At high magnetic field, our result indicates that normal and Andreev reflections become nearly equal through MAR in the quantum Hall regime.
\section{Experimental}

\subsection{Sample fabrication}

\begin{figure}[tbp]
  \begin{center}
    \includegraphics{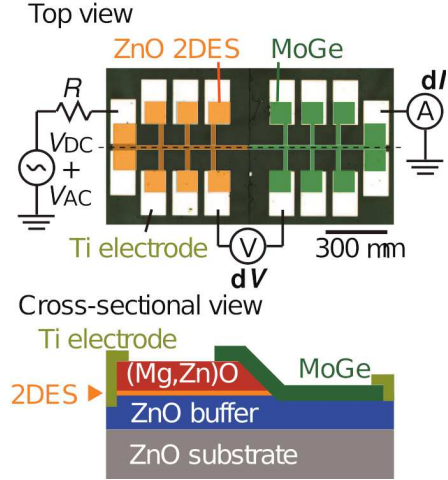}
  \end{center}
  \caption{(color online). Schematics for top and cross-sectional views of the ZnO2DES/MoGe junction. External circuit for the Andreev reflection measurement is also depicted. The series resistance $R$ connected between the voltage source and an electrode is either 10 M$\Omega$ or 100 M$\Omega$ depending on the measurement voltage range.}
\end{figure}

The Mg$_{x}$Zn$_{1-x}$O/ZnO ($x \approx 0.015$) heterostructures were grown by molecular beam epitaxy as described in Ref. \cite{Falson3}. The 2DES/superconductor junction was formed by etching a part of the Mg$_{x}$Zn$_{1-x}$O capping layer by Ar ion milling at an incidence angle of 45$^{\circ}$ followed by deposition of 70-nm thick MoGe with RF magnetron sputtering. Subsequently, the Hall-bar patterns were defined by ion milling for ZnO 2DES and MoGe as shown in Fig 2. The Hall-bar pattern was designed  to enable the measurement of junction properties simultaneous to the transport properties of ZnO 2DES and MoGe. Finally, Ti ohmic electrodes were deposited by electron beam evaporation for ZnO 2DES. The width of the junction ($W$) is 10 $\mu$m.\par

\subsection{Electrical measurement}
The transport properties were measured in a $^{4}$He cryostat down to 2 K and in a dilution refrigerator at a temperature of 0.05 K with lock-in amplifiers at a typical frequency of 7 Hz. For the Andreev reflection measurement, while the alternating voltage ($V_{\mathrm{AC}}$) is superimposed on the bias voltage ($V_{\mathrm{DC}}$), alternating component of the current (d$I$) and the differential voltage including the junction and a part of the ZnO channel were measured by lock-in amplifiers as shown in Fig. 2. In addition, the channel resistances of ZnO 2DES and MoGe were simultaneously measured by extra lock-in amplifiers (not shown in the figure). The differential voltage across the junction (d$V$) was calculated by subtracting the voltage drop in the ZnO channel (d$V_{\mathrm{ZnO}}$) from the measured voltage drop (d$V_{\mathrm{M}}$) as $\mathrm{d}V = \mathrm{d}V_{\mathrm{M}} - \mathrm{d}V_{\mathrm{ZnO}}$. The differential conductance was finally calculated as d$I$/d$V$.\par

\subsection{Modified Blonder-Tinkham-Klapwijk model}

\begin{figure}[tbp]
  \begin{center}
    \includegraphics{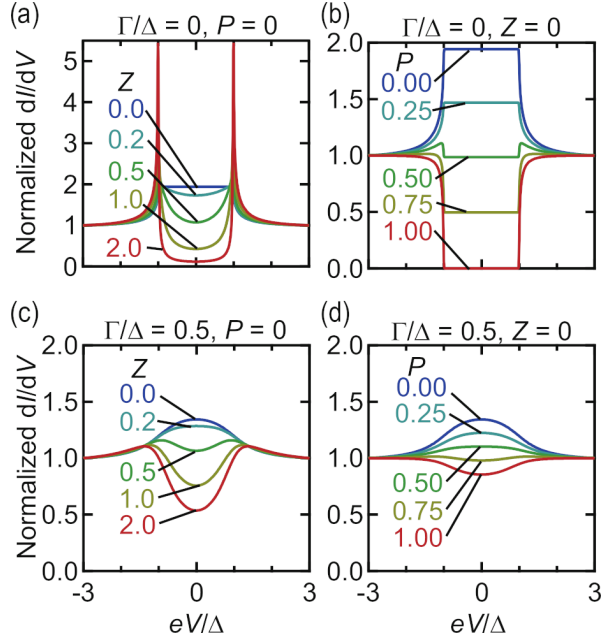}
  \end{center}
  \caption{(color online). Calculated d$I$/d$V$ normalized at $|eV/\Delta| = 3$ as a function of $eV/\Delta$. Interface barrier strength ($Z$) varies in (a), while spin polarization parameter ($P$) varies in (b) assuming zero broadening parameter ($\Gamma/\Delta=0$), meaning infinite quasiparticle lifetime. $Z$ and $P$ dependences are shown in (c) and (d), respectively, assuming a relatively strong quasiparticle scattering of $\Gamma/\Delta=0.5$}
\end{figure}

For analyzing Andreev reflection, we employed the modified Blonder-Tinkham-Klapwijk (BTK) model. In the BTK model, the current for spin-unpolarized and polarized electrons across the interface is expressed as \cite{Blonder1,Plecenik,Raychaudhuri,Auth,Chen}
\begin{equation}
I_{\mathrm{u/p}}\propto\int^{\infty}_{-\infty}\left[f(E-eV)-f(E)\right]\left[1+A_{\mathrm{u/p}}(E)-B_{\mathrm{u/p}}(E)\right]\mathrm{d}E,
\end{equation}
where subscripts u and p denote spin-unpolarized and -polarized current, respectively, $f(E)$ is the Fermi-Dirac distribution function, $E$ is the energy of electrons, $e$ is the elementary electric charge, $V$ is the voltage applied across the junction, $A_{\mathrm{u/p}}(E)$ and $B_{\mathrm{u/p}}(E)$ are the coefficients giving the probability of Andreev and normal reflection, respectively. With taking into account of spin polarization $P$ ($0<P<1$) of the metal (ZnO 2DES in the present case) \cite{deJong,Soulen,Mazin,Deutscher,Ma,Zyuzin,Fisher}, total current is calculated as
\begin{equation}
\label{BTK}
I\propto(1 - P)I_{\mathrm{u}} + PI_{\mathrm{p}}.
\end{equation}
\par

The probabilities of Andreev ($A_{\mathrm{u/p}}$) and normal reflections ($B_{\mathrm{u/p}}$) are calculated by solving Bogoliubov equations with proper boundary conditions as
\begin{eqnarray}
A_{\mathrm{u}}(E) &=& \frac{\sqrt{(\alpha^2+\eta^2)(\beta^2+\eta^2)}}{\gamma^2},\\
B_{\mathrm{u}}(E) &=& Z^{2}\frac{\left[(\alpha-\beta)Z-2\eta\right]^{2}+\left[2\eta Z + (\alpha-\beta)\right]^{2}}{\gamma^2},\\
A_{\mathrm{p}}(E) &=& 0,\\
B_{\mathrm{p}}(E) &=& \frac{\left[Z(\alpha-\beta)-\eta\right]^{2} + (2\eta Z-\beta)^{2}}{\left[Z(\alpha-\beta)+\eta\right]^{2} + (2\eta Z-\alpha)^{2}},
\end{eqnarray}
where $\alpha$, $\beta$, and $\eta$ are defined from Bogoliubov coherence factors ($u_{0}$, $v_{0}$) as
\begin{eqnarray}
u_{0}^{2} &=& \frac{1}{2}\left[1+\frac{\sqrt{(|E|+i\Gamma)^2-\Delta^2}}{|E|+i\Gamma}\right] = \alpha + i\eta,\\
v_{0}^{2} &=& 1-u_{0}^{2} = \beta + i\eta,
\end{eqnarray}
and
\begin{equation}
\gamma^{2} = \left[\alpha + Z^{2}(\alpha - \beta)\right]^{2} + \left[\eta(2Z^{2}+1)\right]^{2}.
\end{equation}

Here, we introduced quasi-particle broadening energy $\Gamma=\hbar/\tau$ ($\hbar$: Planck constant divided by 2$\pi$, $\tau$: quasi-particle lifetime) to describe broadening of Andreev spectra \cite{Dynes}. $Z = V_{0}/\hbar v_{\mathrm{f}}$ is the energy barrier strength parameter between the metal and superconductor for given barrier height $V_{0}$ and Fermi velocity $v_{\mathrm{f}}$ of the metal. Finally, the experimental differential conductance normalized by that at $|V|\gg \Delta/e$ was fitted with the theoretical expression based on Eq. (\ref{BTK}).\par

In order to obtain an insight into the dependence of the d$I$/d$V$ spectrum on the various parameters in modified BTK model, we show calculation results based on Eq. (\ref{BTK}) in Fig. 3. In the ideal case, that is no quasiparticle broadening ($\Gamma/\Delta=0$), no spin polarization ($P=0$), and no energy barrier ($Z=0$), d$I$/d$V$ exhibits nearly twice enhancement within the voltage corresponding to superconducting gap ($|eV|<\Delta$), indicating complete Andreev reflection. By increasing the energy barrier strength at the interface (increasing $Z$), the d$I$/d$V$ spectrum is changed from enhancement to tunneling-type showing sharp peaks indicating quasiparticle density of states at $|eV|=\Delta$ [Fig. 3(a)]. On the other hand, increasing the spin polarization parameter $P$ simply varies the normalized d$I$/d$V$ from 2 to 0 without quasiparticle peaks [Fig. 3(b)]. When nontrivial quasiparticle broadening is turned on ($\Gamma/\Delta=0.5$ for example), all the d$I$/d$V$ spectrum become blurred as shown in Figs. 3(c) and 3(d), which actually better reproduce our experimental results as discussed later.

\section{Results}
\subsection{Basic transport properties of ZnO 2DES and MoGe}

\begin{figure}[tbp]
  \begin{center}
    \includegraphics{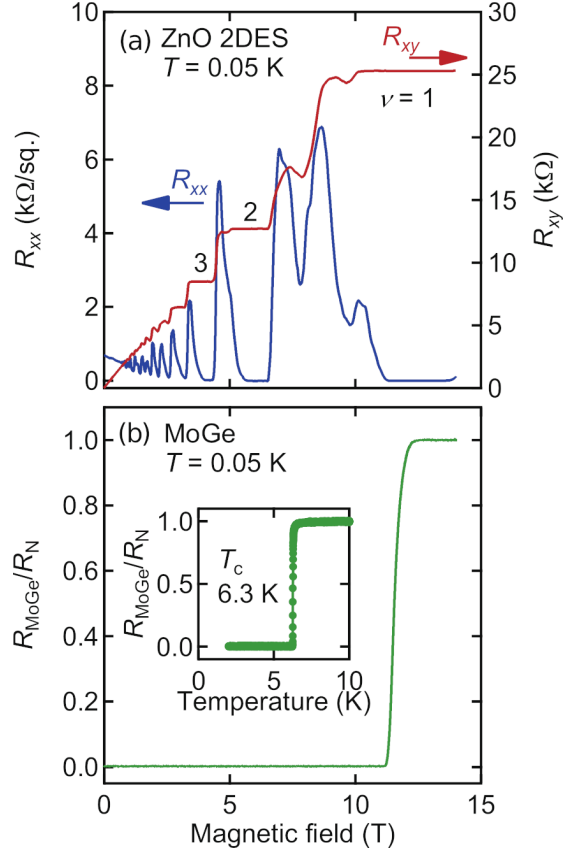}
  \end{center}
  \caption{(color online). (a) Magnetic field dependence of longitudinal resistance $R_{xx}$ and Hall resistance $R_{xy}$ of ZnO 2DES measured at 0.05 K. The carrier density ($n$) and electron mobility obtained from the magnetotransport are $3.0 \times 10^{11}$ cm$^{-2}$ and 310 000 cm$^{2}$ V$^{-1}$ s$^{-1}$, respectively. (b) Magnetic field dependence and temperature dependence (inset) of the resistance ($R_{\mathrm{MoGe}}$) of the MoGe thin film normalized by the normal resistance ($R_{\mathrm{N}} \approx 100$ $\Omega$ )}
\end{figure}

First, we characterize the basic transport properties for ZnO 2DES and MoGe individually as shown in Figs. 4(a) and 4(b). From the low-field magnetotransport, the carrier density and the mobility of the ZnO 2DES used in this study are estimated as $3.0 \times 10^{11}$ cm$^{-2}$ and 310 000 cm$^{2}$ V$^{-1}$ s$^{-1}$, respectively. At 0.05 K, the ZnO 2DES exhibits a number of integer quantum Hall states as evidenced by the vanishing $R_{xx}$ and quantized $R_{xy}$ [Fig. 1(b)]. The superconducting transition temperature ($T_{\textrm{c}}$) and  the upper critical field ($B_{\mathrm{c2}}$) of MoGe are 6.3 K and 11.5 T, respectively [Fig. 4(b)].\par

\subsection{Andreev reflection at $B=0$ T}

\begin{figure}[tbp]
  \begin{center}
    \includegraphics{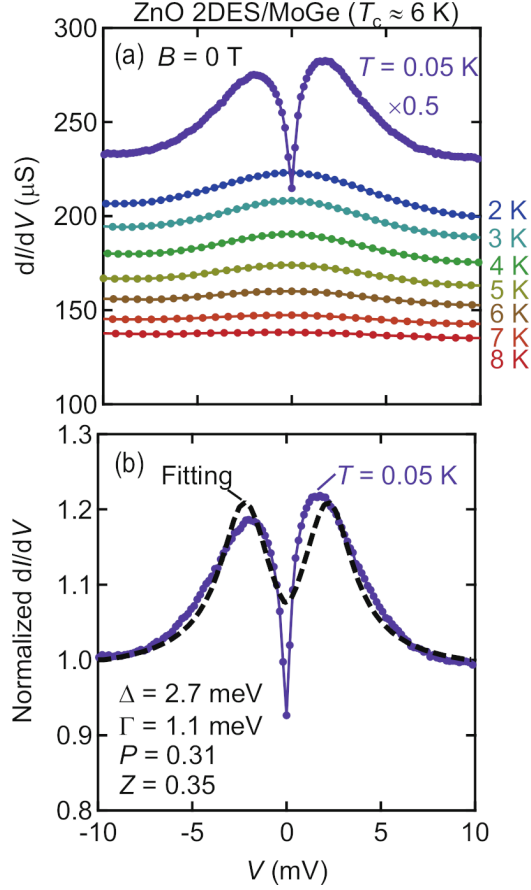}
  \end{center}
  \caption{(color online). (a) Differential conductance of the ZnO2DES/MoGe junction (d$I$/d$V$) as a function of the voltage across the interface ($V$) at several temperatures from 8 K down to 0.05 K. (b) The differential conductance at 0.05 K normalized at $V=-10$ mV. The fit based on Eq. (\ref{BTK}) is also shown together with the fitting parameters.}
\end{figure}

Having the ZnO 2DES and MoGe individually characterized, we measure differential conductance of the junction (d$I$/d$V$) as a function of $V$ at several temperatures as shown in Fig. 5(a). At temperatures above $T_{\mathrm{c}}$, d$I$/d$V$ is almost constant with $V$, indicating an ohmic junction. With decreasing temperature, d$I$/d$V$ gradually increases around $V = 0$ V, indicative of Andreev reflection. At the lowest temperature of $T=0.05$ K, sharp dip appears in addition to the broad enhancement in d$I$/d$V$. Figure 5(b) shows the fitting of the differential conductance at $T=0.05$ K, which is normalized at $V = -10$ mV,  using the modified BTK model with $\Delta$, $\Gamma$, $P$, and $Z$ being free parameters. The modified BTK model overall well fits the experimental data although the sharp dip around $V= 0$ V is not well reproduced. The resultant fitting parameters are $\Delta = 2.7$ meV, $\Gamma = 1.1$ meV, $P = 0.31$, and $Z = 0.35$. The $\Delta$ value is significantly larger than that of bulk value (1.1 meV \cite{Grayl,Bezryadin}) and the broadening factor $\Gamma$ is relatively large. We also obtain finite $P$ although ZnO is nominally nonmagnetic. These unexpected fitting parameters may be due to various nonideal effects frequently existing at semiconductor/superconductor junctions such as damage layer in ZnO 2DES or spatial variation of $\Delta$ near the interface \cite{Neurohr}. In this respect, we limit our discussion within the qualitative dependence of Andreev spectra on magnetic field (filling factor) in the following discussions.

\subsection{Andreev reflection at low magnetic field ($0< B < 2$ T)}

\begin{figure}[tbp]
  \begin{center}
    \includegraphics{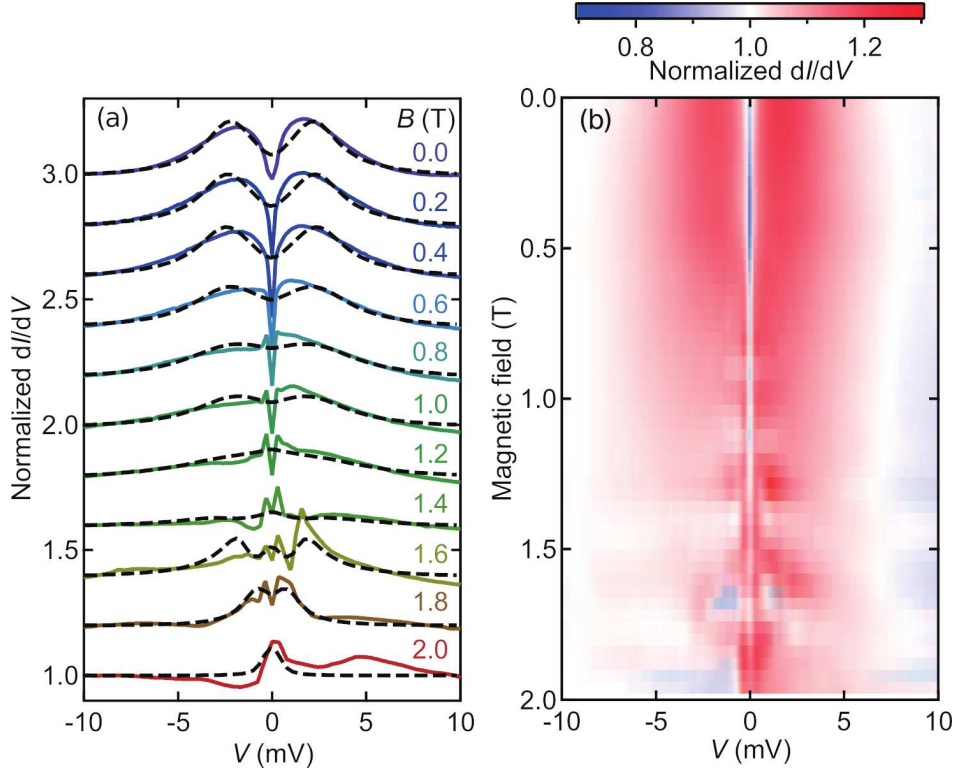}
  \end{center}
  \caption{(color online). (a) Differential conductance d$I$/d$V$ normalized at $V=-10$ mV as a function of $V$ at several magnetic fields below 2 T (solid curves). The dashed curves are fits to the experimental d$I$/d$V$ for negative $V$ region. The curves are shifted vertically for clarity. (b) The color map of the normalized d$I$/d$V$ as functions of magnetic field and $V$.}
\end{figure}

\begin{figure}[tbp]
  \begin{center}
    \includegraphics{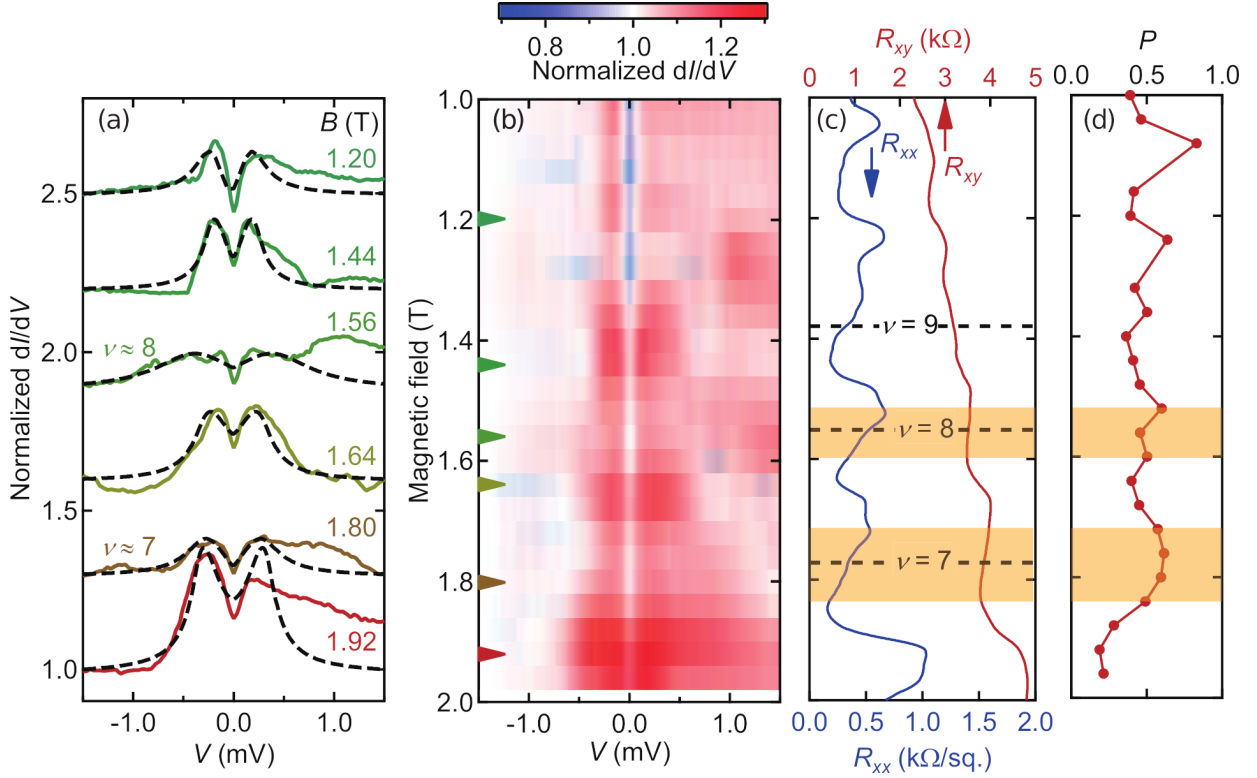}
  \end{center}
  \caption{(color online). (a) Differential conductance d$I$/d$V$ normalized at $V=-1.5$ mV as a function of $V$ at several magnetic fields between 1 T and 2 T(solid curves). The dashed curves are fits to the experimental d$I$/d$V$ for negative $V$ region. The curves are shifted vertically for clarity. (b) The color map of the normalized d$I$/d$V$ as functions of magnetic field and $V$. The triangles indicate the magnetic fields taken in the panel (a). (c) Replot of $R_{xx}$ and $R_{xy}$ for comparison with Andreev spectra in the panel (b). (d) The value of a fitting parameter $P$ as a function of magnetic field. The shaded areas correspond to magnetic field regions of small conductance enhancement shown in the panel (b).}
\end{figure}

Next we investigate Andreev reflection with the application of a small ($B<2$ T) magnetic field. As shown in Fig. 6, The normalized d$I$/d$V$ does not significantly change the spectral shape up to $\sim 1$ T, which exhibits broad conductance enhancement and sharp dip around $V=0$ V. As shown in Figs. 6(a) and 6(b), the dip structure around $V=0$ V becomes sharper with increasing magnetic field toward 1 T compared with that near 0 T. This narrowing means that, under magnetic field, apparent superconducting gap becomes small although the exact reason is not clear at the moment. In order to analyze the detail, we measured d$I$/d$V$ in the narrower voltage range of $-1.5 < V< 1.5$ mV above 1 T with a finer voltage step as shown in Fig. 7. Since, above 1 T, d$I$/d$V$ is almost constant at $|V|>1.5$ mV, the spectra are normalized at $V=-1.5$ mV. Here $R_{xx}$ and $R_{xy}$ of ZnO 2DES are also replotted in the panel (c) to be compared with Adnreev spectra in the panels (a) and (b). At some magnetic fields, d$I$/d$V$ shows nontrivial asymmetry with $V$ and therefore we fit only the negative $V$ region as shown in Fig. 7(a). Overall, conductance enhancement is observed around $|V|<0.5$ mV together with a dip around $V=0$ V originating from the effect of $Z$. However, the effect of $Z$ is less dominant ($Z\lessapprox0.3$) at magnetic fields higher than 1.4 T, and is not focused on here. If we closely look at the magnetic field dependence, we find that the height of the normalized d$I$/d$V$ appears to oscillate. For instance, as shown in Fig. 7(a), the normalized d$I$/d$V$ at $B = 1.56$ T ($\nu\approx8$) and $1.80$ T ($\nu\approx7$) are clearly smaller than those at other magnetic fields, where $\nu$ is defined as $nh/eB$ ($h$: Planck constant). The shaded area in the panels (c) and (d) indicate the range of relatively small d$I$/d$V$. In terms of modified BTK formula, as discussed in Fig. 3, the conductance enhancement or reduction is nominally reflected in the spin polarization parameter $P$. The $P$ value is plotted as a function of magnetic field in Fig. 7(d). Comparing the Andreev spectra in Fig. 7(b) and the magnetotransport of ZnO 2DES in Fig. 7(c), relatively high $P$ regions roughly corresponds to the developing quantum Hall plateau. The interpretation of the behavior of $P$ will be discussed later.

\subsection{Andreev reflection in the quantum Hall region ($ B > 2$ T, $\nu<6.1$)}

\begin{figure}[tbp]
  \begin{center}
    \includegraphics{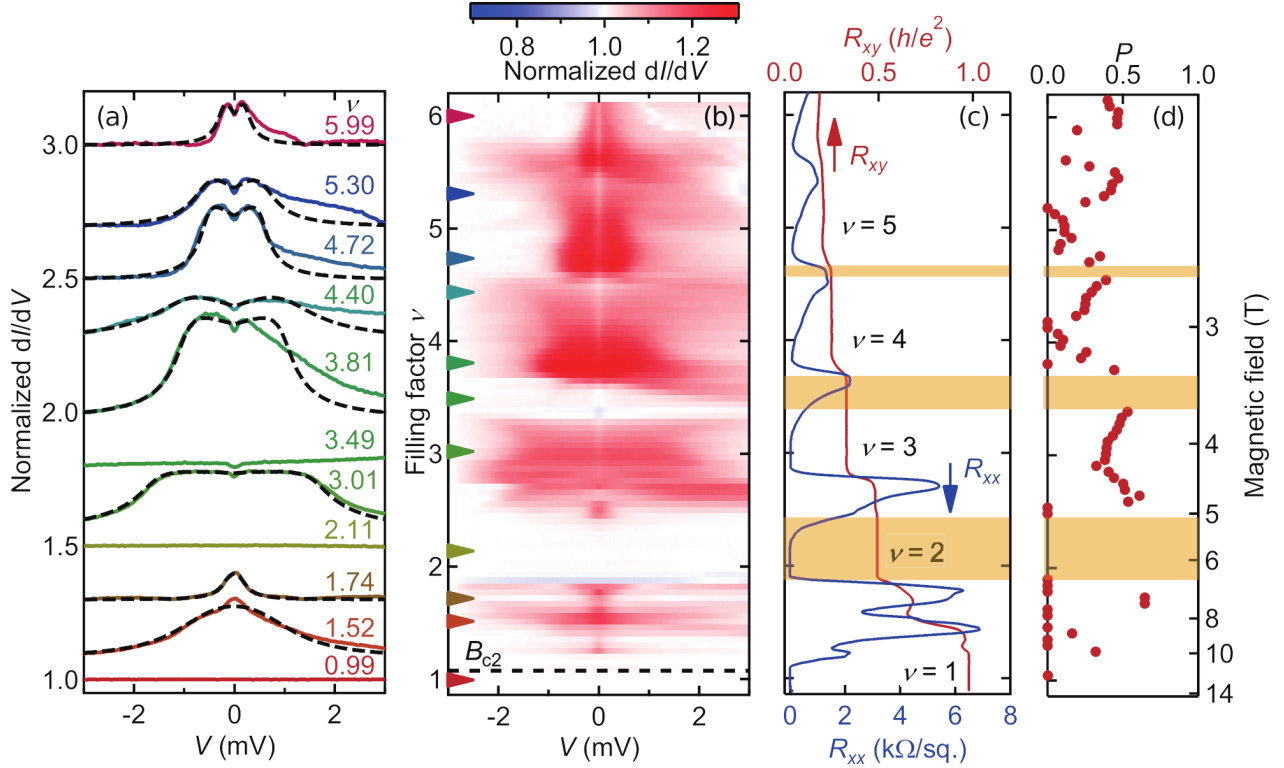}
  \end{center}
  \caption{(color online). (a) Differential conductance d$I$/d$V$ normalized at $V=-3.0$ mV as a function of $V$ at several filling factors $\nu<6.1$ corresponding to the magnetic field range of $B>2$ T (solid curves). The dashed curves are fits to the experimental d$I$/d$V$ for negative $V$ region. The curves are shifted vertically for clarity. (b) The color map of the normalized d$I$/d$V$ as functions of filling factor and $V$. The triangles indicate the filling factors taken in the panel (a). (c) Replot of $R_{xx}$ and $R_{xy}$ as a function of $\nu$ for comparison with Andreev spectra in the panel (b). (d) The value of a fitting parameter $P$ as a function of filling factor. The shaded areas correspond to magnetic field regions of small conductance enhancement shown in the panel (b).}
\end{figure}

The effect of the Landau level quantization on Andreev reflection is more pronounced at higher magnetic fields in the well developed quantum Hall regime. Figure 8 shows the Andreev spectra above 2 T, corresponding to a filling factor range of $\nu<6.1$. Here, d$I$/d$V$ is normalized at $V=-3$ mV as the spectra tend to saturate at larger voltages. Most of the d$I$/d$V$ spectra are well fitted although the spectral shape is sometimes asymmetric with $V$. We find that the strength of the conductance enhancement [red regions in Fig. 8(b)] periodically varies with the filling factor, indicating that Andreev spectra indeed reflect the quantum Hall edge states. As in the case of low magnetic field in Fig. 7, d$I$/d$V$  is suppressed in some regions of filling factors as shown in Figs. 8(a) and 8(b). Particularly around $\nu=2$, d$I$/d$V$ is almost constant with $V$ as shown in Fig. 8(a), and reliable fitting is not feasible. Instead, we compare the experimental d$I$/d$V$ with modified BTK calculation in Fig. 3. Since experimental d$I$/d$V$ spectra around $\nu=2$ do not show quasiparticle peaks, the effect of the $Z$ parameter is not dominant and the comparison is made between the experimental data in Fig. 8(a) and calculation in Fig. 3(d), where $P$ parameter is varied, indicating that $P\approx 0.5$--$0.7$ relatively well reproduce almost constant d$I$/d$V$.

\section{Discussions}
So far, we observe that the magnetic field (filling factor) dependence of Andreev reflection exhibits periodic modulation in d$I$/d$V$, indicating that Laudau level quantization plays an important role. In terms of fitting by Eq. (\ref{BTK}), the conductance enhancement or reduction is dominantly reflected in the $P$ value as seen in Fig. 3. Conventionally, $P$ is interpreted as degree of spin polarization in the metal or semiconductor because spin-polarized density of states prohibits Andreev reflection and interface conductance is suppressed towards zero at full spin polarization of $P=1$. However, in a more general sense, $P$ reflects the probability of normal reflection [Fig. 1(b)] with respect to Andreev reflection [Fig. 1(a)] \cite{deJong,Soulen,Mazin}. In the quantum Hall regime, $P$ may effectively be interpreted as the ratio of normal reflection with respect to  Andreev reflection in the MAR process shown in Fig. 1(c). Then, the flat d$I$/d$V$ [e.g. at $\nu = 3.49$ or at $\nu = 2.11$ in Fig. 8(a)] means nearly equal probability of Andreev and normal reflections. This is consistent with a simple idea that, when Andreev reflection probability is between 0 and 1 in a single reflection process, total Andreev reflection probability becomes 0.5 after MAR process. In our experiment, since the cyclotron radius ($R_{\mathrm{C}} = 13$ nm at 5 T) is much smaller than the width of the interface ($W = 10$ $\mu$m), this picture may be valid, which has been more precisely calculated in several reports. In Ref. \cite{Chtchelkatchev}, where MAR is treated as successive single Andreev or normal reflections in the high $\nu$ region with taking into account the interface disorder (roughtness), conductance oscillations periodic in filling factor $\nu$ are theoretically predicted, which is consistent with our data  in Figs. 7 and 8. However, in Ref.  \cite{Giazotto}, where electron and hole skipping cyclotron orbits at the 2DES/superconductor interface are theoretically calculated at low $\nu$ based on Bogoliubov--de Gennes equation, conductance enhancement appears every two indices in $\nu$ due to Zeeman splitting, which is not the case in our experiment. One reason would be that the simple Zeeman splitting picture may not be valid in the case of ZnO 2DES owing to strong correlation \cite{Maryenko2}.\par

Finally, we compare our result with recent experiments in other materials systems. As mentioned in the introduction, since most of semiconductors form Shottky barrier, InAs and graphene may be suitable materials for comparison. Most studies use Josephson structures comprised of superconductor-2DES-superconductor with the interface width and channel length of $\sim 1$ $\mu$m. They observed supercurrent at $B=0$ T as well as conductance enhancement at quantum Hall plateaux in some cases \cite{Komatsu,Rickhaus,Amet}. Single superconductor/2DES junctions have also been investigated, which showed oscillating d$I$/d$V$ periodic in $\nu$, similar to our study \cite{Park,Matsuo}. In Ref. \cite{Lee}, on the other hand, full conversion from an electron to a hole occurs only when the width of  the superconductor is less than the coherence length of the Cooper pair ($\approx 50$ nm). A particularly important question for realizing exotic interface states is whether complete Andreev process is necessary or partial conversion from an electron to a hole is sufficient, which remains to be investigated in the future.

\section{Conclusions}
In this study, we have investigated electrical properties across the interface between two-dimensional electron system of ZnO and MoGe superconductor particularly in the quantum Hall states utilizing Andreev reflection spectroscopy. At low temperature, conductance enhancement indicating Andreev reflection is observed. With applying magnetic field, the conductance exhibits oscillations periodic in filling factor, signaling a proximity effect between the quantum Hall edge states and Cooper pairs. The Andreev spectra are analyzed based on the modified Blonder-Tinkham-Klapwijk equation. The result indicates that, in the quantum Hall regime, electrons incident on an edge of the 2DES/superconductor interface are partially converted to holes at the other edge of the interface through multiple Andreev reflection. However, conductance enhancement is suppressed in the regions of quantum Hall plateaux, giving rise to a $P$ parameter of $\approx 0.5$--$0.7$, which can be interpreted as about a half probability of Andreev reflection with the rest of normal reflection through multiple Andreev reflection along the interface. This is probably due to long interface of 10 $\mu$m used in  this study, and phase coherence between the incident electrons and outgoing holes are randomized. Regarding the non-abelian particles predicted to appear at the interface between helical or chiral edges and a superconductor, our study shows that two-dimensional electron system in ZnO may be one of the suitable systems because of the good proximity with a superconductor and the high electron mobility of the 2DES.

\section{Acknowledgment}
We appreciate T. Tamegai and M. Kawamura for valuable suggestions. This research was supported by JST, PRESTO Grant Number JPMJPR1763 and JST, CREST Grant Number JPMJCR16F1, Japan, as well as by Mitsubishi Foundation (Y.K.). This work was carried out by joint research of the Cryogenic Research Center, the University of Tokyo.

\end{document}